\documentclass[a4paper,UKenglish,cleveref,autoref,thm-restate]{lipics-v2021}

\pdfoutput=1 
\hideLIPIcs  

\usepackage{booktabs}
\usepackage{graphicx}
\usepackage{subcaption}
\usepackage{algorithm2e}
\usepackage{siunitx}

\nolinenumbers
\graphicspath{{images/}}
\usepackage{svg}

\title{Benchmarking and Engineering Data Structures for Spherical Range Queries} 

\author{Thomas Bläsius}
{Karlsruhe Institute of Technology (KIT), Karlsruhe, Germany}
{thomas.blaesius@kit.edu}
{https://orcid.org/0000-0003-2450-744X}
{}

\author{Jean-Pierre von der Heydt}
{Karlsruhe Institute of Technology (KIT), Karlsruhe, Germany}
{heydt@kit.edu}
{https://orcid.org/0009-0000-3852-350X}
{This work was supported by funding from the pilot program Core--Informatics of the Helmholtz Association (HGF).}

\author{Tobias Kempf}
{Karlsruhe Institute of Technology (KIT), Karlsruhe, Germany}
{tobias.kempf@student.kit.edu}
{https://orcid.org/0009-0001-3327-4652}
{}

\author{Dennis Kobert}
{Karlsruhe Institute of Technology (KIT), Karlsruhe, Germany}
{dennis.kobert@student.kit.edu}
{https://orcid.org/0009-0005-6964-495X}
{}
\author{Nikolai Maas}
{Karlsruhe Institute of Technology (KIT), Karlsruhe, Germany}
{nikolai.maas@kit.edu}
{https://orcid.org/0009-0002-6959-417X}
{}

\authorrunning{T. Bläsius, J.-P. von der Heydt, T. Kempf, D. Kobert and N. Maas} 

\Copyright{Thomas Bläsius, Jean-Pierre von der Heydt, Tobias Kempf, Dennis Kobert and Nikolai Maas} 



\category{} 

\relatedversion{} 

\supplement{}
\supplementdetails[subcategory={}, cite={}, swhid={}]{Source Code}{https://github.com/wembed-pdf/sprk}
\supplementdetails[subcategory={}, cite={}, swhid={}]{Benchmarking Code}{https://github.com/wembed-pdf/rembed}
\supplementdetails[subcategory={}, cite={}, swhid={}]{Embedding Dataset}{https://doi.org/10.5281/zenodo.21243483}

\begin{document}

\maketitle

\newcommand{\CC}{C\nolinebreak\hspace{-.05em}\raisebox{.4ex}{\tiny\bf +}\nolinebreak\hspace{-.10em}\raisebox{.4ex}{\tiny\bf +}}

\begin{abstract}

	Spherical range queries are a fundamental primitive for working with spatial data.
	Many spatial data structures have been developed to answer these queries, but choosing the optimal one for a specific application is a difficult task.
	This is because theoretical worst-case bounds are often overly pessimistic, and existing average-case analyses are rather restricted and hard to compare.

	We address this problem with two main contributions.
	First, we present a comprehensive evaluation of state-of-the-art spatial indices across a diverse set of benchmarks.
	This includes a new benchmark based on graph embeddings alongside multiple real-world datasets from the literature. 
	Our benchmark covers instances scaling up to $\SI{10}{M}$ points and ranging between \num{2} and \num{960} dimensions.
	
	Second, we introduce the Sorted-Projection Radius KD-tree (SPRK-tree), a high-performance KD-tree variant.
	The SPRK-tree combines aggressive subtree pruning via radius reduction, sorted projection-based leaf nodes, and careful implementation optimizations. 
	It consistently achieves the fastest query times in almost all benchmarks,
	and ranks second in the few remaining cases.
\end{abstract}

\newpage

\section{Introduction}\label{sec:intro}

Spherical range queries are a fundamental primitive for working with spatial data.
Given a set of $n$ points in $d$-dimensional Euclidean space,
the objective is to retrieve all points that lie within the specified radius of a query point.
In low-dimensional spaces, spatial indices significantly accelerate these queries by pruning large portions of the search space.
However, in higher dimensions a brute-force approach that checks the distance to every point 
is often the only viable option due to the curse of dimensionality~\cite{indykApproximate1998}. 

One application of spatial indices is the computation of graph embeddings~\cite{hyperbolicPlaneEmbedding, blasiusWeighted2025, forceatlas2, tsne}, which play an important role in machine learning tasks such as community detection, link prediction, and node classification~\cite{communityDetection, node2vec, linkPrediction, deepwalk, verse}.
These methods map the vertices of a graph into a continuous space such that spatial distance reflects structural similarity~\cite{makarovSurvey2021}.
This is often achieved by force-directed optimization, which applies attracting forces to adjacent vertices that are too distant from each other and repelling forces to non-adjacent vertices that are too close.
The attracting forces can be computed in linear time $O(m)$ by iterating over all $m$ edges.
However, computing repelling forces requires identifying all non-adjacent vertices within a specific distance of each other.
A naive brute-force implementation would check all $O(n^2)$ pairs of vertices, which is particularly prohibitive for large sparse graphs.
Alternatively, the close vertex pairs can be identified by performing a spherical range query for each vertex, making the performance of the embedding framework heavily dependent on the efficiency of the underlying spatial index.

Beyond graph embedding, range queries also play a crucial role in numerous other domains
such as particle-based simulations, database systems, and computer graphics~\cite{gaedeMultidimensionalAccessMethods1998, ihmsenSPHFluidsComputer2014, jensenGlobalIlluminationUsing1996}.
For a comprehensive overview of applications, see the book by Samet~\cite{sametFoundations2006}.
To support these applications, numerous spatial indices have been developed; see \cref{tab:spatial-complexity} for a list.
The theoretical worst-case bounds also shown in \cref{tab:spatial-complexity} give the impression that these spatial indices do not yield significant improvements over the brute-force approach.
However, this is highly misleading and comes from the fact that the worst-case is often too pessimistic.
In practice, the different spatial indices are often successful in pruning large chunks of the search space.
In the following, we give a brief overview of the different spatial indices and broadly categorize them by how they organize and prune the search space.

The first category of spatial indices is based on partitioning the geometric space.
Uniform grids partition space into fixed-size cells and insert the points into them.
Queries are answered by a lookup of the corresponding cell and inspecting it and adjacent cells.
These yield especially fast lookups in low dimensions when the points are evenly distributed.
However, they lack adaptability and in datasets with heterogeneous density, dense cells often become a bottleneck.
Orthtrees (also known as Quadtrees in 2D and Octrees in 3D)~\cite{sametOverviewQuadtreesOctrees1988} address this by recursively dividing the space only where necessary.
While more flexible than grids, they share the same limitation, where each split generates $2^d$ children, which is infeasible in higher dimensions.

Unlike space partitioning methods, data partitioning methods divide the point set itself.
KD-trees~\cite{bentleyMultidimensionalBinarySearch1975, friedmanAlgorithmFindingBest1977} split the point set along axis-aligned hyperplanes, to construct a balanced tree.
To answer a query, the tree is traversed recursively, pruning subtrees that cannot contain query results.
R-trees~\cite{guttmanRtreesDynamicIndex1984} group points into bounding rectangles, VP-trees~\cite{brinNeighbor1995, yianilosDataStructuresAlgorithms1993} partition by distance to a vantage point, and Balltrees~\cite{omohundroFiveBalltreeConstruction1989} use bounding hyperspheres.

The SNN structure of Chen and Güttel~\cite{chenFastExactFixedradius2024} avoids hierarchical traversal entirely.
It uses the singular value decomposition (SVD) to identify the direction of maximum variance in the dataset.
Points are projected onto this axis and sorted.
The query algorithm uses a binary search to identify a contiguous range of candidate points that is then scanned exhaustively.

This shows that there are plenty of spatial indices to choose from.
As the theoretical worst-case bounds do not provide strong guidance, choosing the right data structure for the application at hand, e.g., graph embedding, can be a daunting task.
While there are some average-case results that give guarantees in certain situations~\cite{bentlyDataStructuresRange1979}, they are rather restricted and sometimes difficult to compare.
Moreover, the performance depends on the data distribution of the specific application as well as implementation details.
Consequently, an empirical evaluation is essential to choose the best data structure.
The only empirical study in this direction we are aware of is by Lawson, Gropp, and Lofstead~\cite{Explor_Spatial_Index_Accel_Featur_Retriev_HPC-Lawson22}, whose comparison is tailored for the HPC-context, focusing on orthogonal range queries in 3-dimensional space 
in a distributed setting.
Additionally, some libraries, e.g., the KD-tree library Kiddo~\cite{donnellysdd2026}, include benchmarks.
However, to the best of our knowledge, there has been no extensive scientific comparison of spatial indices on a diverse set of inputs with varying dimensions yet.

\begin{table}
	\centering
	\caption{Theoretical complexity of spatial data structures. 
		\textit{Variables:} $n$ = number of points; $d$~=~number of dimensions; $k$ = output size; $g$ = grid resolution (number of cells per dimension)}\label{tab:spatial-complexity}
	\begin{tabular}{@{}llll@{}}
		\toprule
		\textbf{Structure}                                                                       & \textbf{Construction} & \textbf{Space}        & \textbf{Query (Worst-case)} \\
		\midrule
		Brute-Force                                                                              & $O(1)$                & $O(nd)$               & $O(nd)$                     \\
		Uniform Grid\footnotemark                                                                             & $O(nd + g^d)$         & $O(nd + g^d)$         & $O(nd + 2^d)$               \\
		Orthtree~\cite{sametOverviewQuadtreesOctrees1988}                               		 & $O(n \cdot 2^d)$      & $O(n \cdot 2^d)$      & $O(n\cdot 2^d)$               \\
		KD-tree~\cite{bentleyMultidimensionalBinarySearch1975, friedmanAlgorithmFindingBest1977} & $O(dn \log n)$        & $O(nd)$               & $O(d \cdot n^{1-1/d} + dk)$ \\
		R-tree~\cite{guttmanRtreesDynamicIndex1984}                                              & $O(dn \log n)$        & $O(nd)$               & $O(nd)$                     \\
		VP-tree~\cite{yianilosDataStructuresAlgorithms1993, brinNeighbor1995}                    & $O(dn \log n)$        & $O(nd)$               & $O(nd)$                           \\
		Balltree~\cite{omohundroFiveBalltreeConstruction1989}                                    & $O(dn \log n)$        & $O(nd)$               & $O(nd)$                    \\
		SNN                                                                                      & $O(dn \log n + nd^2)$ & $O(nd)$               & $O(nd)$                     \\
		\bottomrule
	\end{tabular}
\end{table}  
\footnotetext{Grid: for the stated query time, the query radius must be known at construction time.}

\subsection{Our Contribution}
\label{sec:our-contribution}

We have two main contributions.
First, we address the need for an empirical comparison by providing a comprehensive evaluation of different state-of-the-art spatial indices.
Secondly, we provide a new KD-tree variant that we call \emph{Sorted-Projection Radius KD-tree} (SPRK-tree; pronounced ``spark''), which outperforms the competition on almost all benchmarks.
In the following, we discuss these two contributions in more detail; starting with the SPRK-tree.

\subparagraph{SPRK-Tree.}

\makeatletter
\pretocmd{\subparagraph}{\refstepcounter{subsubsection}}{}{}
\makeatother

At its core, the SPRK-tree is a KD-tree variant that recursively partitions the data points 
until the number of points falls below a certain predefined bucket size for the leaves; see \cref{sec:kd-tree}.
In the leaves, we use a variant of SNN~\cite{chenFastExactFixedradius2024}, i.e., we sort the remaining points with respect to a projection line;
see \cref{sec:snn-buckets}.
For the query, we utilize pruning techniques, which reduce the query radius incrementally while traversing the hierarchy~\cite{aryaAlgorithmsFastVector1993, friedmanAlgorithmFindingBest1977, CGAL}.
This allows more aggressive pruning of subtrees, while still ensuring that no points in the original query ball are omitted; see \cref{sec:radius-reduction}.
For high-dimensional data sets, we initially rotate the data based on a singular value decomposition of sampled points, such that the first splits done by the KD-tree separate along directions with high spread; see \cref{sec:svd}.
Our Rust implementation features SIMD vectorization and cache-aware memory layout (see \cref{sec:perf_engineering}), and provides bindings for \CC{} and Python.

\subparagraph{Experimental Evaluation.}

Our evaluation uses a large and diverse set of benchmarks, including a newly created benchmark set.
Our new set comes from the graph embedding application discussed above.
It features over \num{240} embedding instances with up to \SI{1}{M} points in dimensions~$\{2,3,\ldots,16,32\}$.
Beyond that, we also consider points distributed uniformly at random and benchmarks from the literature, including data sets from clustering, point of interest queries, and a high-dimensional real-world data set with point sets of up to \num{960} dimensions.
Besides our SPRK-tree, we evaluate \num{10} other implementations as well as a brute-force approach. 
The evaluated data structures are diverse (Quadtree, uniform grid, Balltree, VP-tree, R-tree, KD-tree, SNN) and include state-of-the-art implementations of KD-trees in multiple programming languages.
Our findings can be summarized as follows.
\begin{itemize}
\item For the embedding benchmark, KD-trees are the method of choice.
  While the KD-tree libraries Kiddo~\cite{donnellysdd2026} and Neighbourhood~\cite{hominusneighbourhood2025} already perform well, our SPRK-tree consistently outperforms them.
  For sufficiently high dimensions, brute-force eventually becomes the best option, yet our SPRK-tree still outperforms brute-force by an order of magnitude for 16 dimensions.
\item For randomly distributed points, tree-based methods perform well for low dimensions but degrade for higher dimensions.
  This is contrasted by SNN and brute-force, which perform much worse on lower dimensions but outperform the tree-based methods for dimension \num{32}.
  The exception to this is our SPRK-tree.
  While being the best tree-based method for low dimensions, it is also (slightly) faster than SNN and brute-force for dimension \num{32}, outperforming the other tree-based methods by an order of magnitude.
\item While we developed the SPRK-tree with the embedding application in mind, it still outperforms the competitors on most benchmarks from other applications.
  The only exception is the benchmark of fixed-radius search in (very) high dimensions, where the SPRK-tree is usually second after SNN or brute-force, but still stays competitive.
\end{itemize}
Beyond the extensive comparison with other data structures, we provide an empirical analysis of how our design and optimization choices contribute to the performance of the SPRK-tree.

\section{The SPRK-tree}\label{sec:SPRK-tree}

In this section we introduce the SPRK-tree, which improves upon
the standard KD-tree through three primary mechanisms:
aggressive pruning of subtrees using radius reduction, the use of SNN-inspired sorted leaf buckets,
and an SVD-based rotation to improve performance on high-dimensional data.
We begin with a brief description of the standard KD-tree, followed by a detailed explanation of our improvements.
We conclude by describing implementation optimizations that further boost the performance of the SPRK-tree. 

\subsection{KD-tree}\label{sec:kd-tree}

\begin{figure}[tb]
	\centering
	\begin{minipage}[t]{0.47\textwidth}
		\centering
		\includegraphics[page=3]{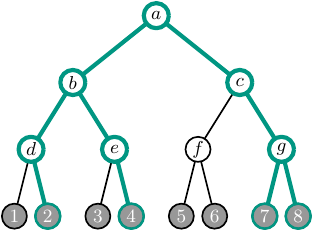}
		\subcaption{Nodes of a KD-tree. Internal nodes are labeled from $a$ to $g$.
		Leaf nodes are numbered and indicated by a small gray circle.
		The thick green edges represent the nodes visited during the query of $q$.
		}\label{fig:kd_tree1}
	\end{minipage}
	\hfill
	\begin{minipage}[t]{0.47\textwidth}
		\centering
		\includegraphics[page=4]{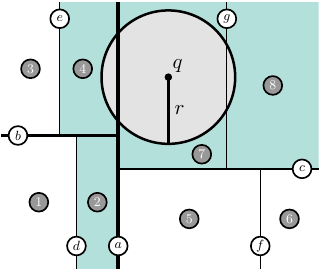}
		\subcaption{Space partitioning of a $2$-dimensional KD-tree and query $q$.
		The lines (subset of hyperplanes) are assigned the label of their inner node.
		The green shaded region indicates the visited leaf nodes.}\label{fig:kd_tree2}
	\end{minipage}
	\caption{Illustration of a $2$-dimensional KD-tree and the nodes visited by a query $q$. 
	 }\label{fig:kd_tree}
\end{figure}

The \emph{KD-tree}~\cite{bentleyMultidimensionalBinarySearch1975, friedmanAlgorithmFindingBest1977}
is a spatial index that organizes a point set $P\subseteq\mathbb{R}^d$ for range queries.
It chooses an axis-aligned hyperplane to split the whole space into two half-spaces 
such that both half-spaces contain the same number of points.
This procedure is applied recursively, yielding a binary tree, where each \emph{internal node} corresponds to a hyperplane and its two children correspond to the half-spaces.
When the number of points falls below a predefined bucket size, the recursion stops and creates a \emph{leaf node}.
Each node is associated with a \emph{region}, 
namely the intersection of all half-spaces on the path from the \emph{root} to the node.
With this, the leaves form a partition of the whole space $\mathbb{R}^d$;
for an example see \cref{fig:kd_tree}. 

A degree of freedom to fill in is the choice of the axis-aligned hyperplane in each step.
We achieve this by cycling through the dimensions in round-robin order 
and placing the hyperplane at the median coordinate of the current points along that dimension.
This halves the point set at every split and produces a tree of logarithmic depth.

A \emph{spherical range query} with query point $q\in\mathbb{R}^d$ and radius $r$ asks for all points in $P$ within distance  $r$ of $q$.
The KD-tree resolves this by recursively traversing the tree and pruning subtrees that cannot contain query results.
Specifically, at an internal node, a child subtree is pruned if the query sphere does not intersect the child's half-space.
Upon reaching a leaf, every point in the leaf is checked against the query radius by an exhaustive search.

\subsection{Radius Reduction}\label{sec:radius-reduction}

Note that the default pruning of a KD-tree does visit unnecessary regions, e.g., Region~$2$ in \cref{fig:kd_tree} despite the fact that Region~$2$ does not intersect the query ball.
This can be prevented by a more informed pruning procedure~\cite{aryaAlgorithmsFastVector1993, friedmanAlgorithmFindingBest1977, CGAL}.
For this, consider the example in \cref{fig:sprk1} (which is the same as in \cref{fig:kd_tree}), and let $p$ be the point of Region~$2$ closest to the query point $q$.
Let $\delta_1$ and $\delta_2$ be the distance between $p$ and $q$ in dimensions $1$ and $2$, respectively.
If we had computed $\delta_1$ and $\delta_2$, we could have pruned early by observing that $\|(\delta_1, \delta_2)\| > r$.
To compute $\delta_1$ and $\delta_2$, note that they are also the distances of $q$ to the hyperplanes $a$ and $b$, respectively, which both separate $q$ from Region~$2$.
This is not a coincidence: for any given region, we can compute the corresponding $\delta_1$ and $\delta_2$ by going up the tree; for each dimension taking the maximum distance between $q$ and a hyperplane (of that dimension) separating $q$ from the region.

Similar to~\cite{aryaAlgorithmsFastVector1993, CGAL}, we do not recompute all $\delta$ values for each node of the tree individually but keep track of a vector $\Delta = (\delta_1, \dots, \delta_d)$ while traversing the tree.
This allows us to update $\|\Delta\|$ using a constant number of arithmetic operations for the distance check.
We start with $\delta_i = 0$ for all $i \in [d]$.
Now let $u$ be a node in the tree with corresponding hyperplane $H_u$.
Consider traversing from $u$ to a child $v$ such that $H_u$ separates the query point $q$ from the region corresponding to $v$.
Then we update $\delta_i$ to the distance from the query point $q$ to the hyperplane $H_u$, where $i$ is the dimension of $H_u$.

\subsection{SNN Buckets}\label{sec:snn-buckets}

\begin{figure}[tb]
	\centering
	\begin{minipage}[t]{0.47\textwidth}
		\centering
		\includegraphics[page=5]{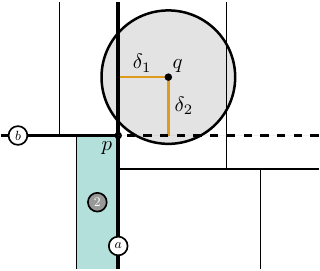}
		\subcaption{While a normal KD-tree would visit Leaf~$2$, 
		the SPRK-tree can prune it since $\|p-q\| = \|(\delta_1, \delta_2)\|$ is larger than the query radius.}\label{fig:sprk1}
	\end{minipage}
	\hfill
	\begin{minipage}[t]{0.47\textwidth}
		\centering
		\includegraphics[page=7]{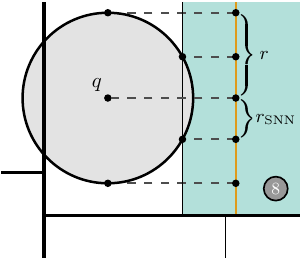}
		\subcaption{Leaf node with an SNN line (orange). 
		The search range can be reduced based on the intersection of the query ball with the region of Leaf~$8$.}\label{fig:sprk2}
	\end{minipage}
	\caption{
		Illustration of the search space pruning in the SPRK-tree.
	}\label{fig:sprk}
\end{figure}

In a regular KD-tree, the points in a leaf node are filtered by a linear scan.
The SPRK-tree replaces this with an SNN data structure~\cite{chenFastExactFixedradius2024} for each leaf.
The SNN structure is constructed by projecting all the points of the leaf onto an axis-aligned line and sorting along its dimension; see Region~$8$ in \cref{fig:sprk2}.
Instead of now scanning all points in the leaf, we can prune some points based on the projection.
We can do this by projecting the query point $q$ to the SNN line and then finding all points within distance $r$ in both directions from $q$ on the line.
This can be done with a binary search followed by a linear scan along the order.
Note that the resulting set of points has to be filtered for points actually in the query ball.

Observe in \cref{fig:sprk2} that the intersection of the query ball with Region~$8$ gives rise to an improvement over this.
In the direction of the SNN line, this intersection spans only a smaller range of size $2\cdot r_{\text{SNN}}$ than the range of $2\cdot r$ queried above.
To compute this smaller radius $r_{\text{SNN}}$, we can utilize the distance vector $\Delta$ from the previous section.
To make this precise, let the dimension of the SNN line be, without loss of generality, the first dimension.
Then $r_{\text{SNN}}$ is the maximum value such that $(r_{\text{SNN}}, \delta_2, \dots, \delta_d)$ has length at most $r$.

Furthermore, to avoid the binary search, the SPRK-tree computes a lookup table for the sorted coordinates of the leaf nodes by discretizing the coordinate range into fixed-width bins. 
Pseudocode for the query procedure of a node is given in \cref{sec:sprk_query}.

\subsection{Singular Value Decomposition}\label{sec:svd}

In many scenarios, high-dimensional data often resides within a lower-dimensional subspace.
The SPRK-tree exploits this property by rotating the data so that the $d$ principal components with the largest variance align in descending order with the coordinate axes.
Because the tree splits dimensions in a round-robin sequence, this rotation ensures that the earliest partitions occur along the directions of maximum spread, enabling more efficient pruning near the root.
The rotation is computed via the SVD of a sampled subset of up to \SI{10}{k} points of $P$, which ensures that the SVD computation does not dominate the overall construction cost.

However, this transformation introduces a slight overhead during the query phase, as every subsequent query requires an initial rotation into the new coordinate space.
Consequently, the SPRK-tree only applies this rotation for datasets with more than \num{16} dimensions. 
In these high-dimensional cases, the tree depth is often smaller than $d$, meaning a standard traversal would never cycle through all dimensions. 
The SVD rotation guarantees that the limited number of splits are used to separate the data along the most informative dimensions.

\subsection{Performance Engineering}\label{sec:perf_engineering}
Beyond algorithmic improvements, we engineered our implementation to utilize the full capabilities of modern hardware. 
In the following, we describe the most important optimizations.

\subparagraph{Distance Computations.}

During a query, the spatial index spends a large part of the time computing distances between the query point and candidate points in the leaf nodes. 
A naive SIMD (Single Instruction, Multiple Data~\cite{flynnComputer1972}) implementation might attempt to accelerate this by computing the differences across the $d$ dimensions of a single point simultaneously. 
However, this can be wasteful since the dimensionality $d$ varies per dataset and does not necessarily align with the fixed width of hardware SIMD registers.

To resolve this, we vectorize across points rather than dimensions. 
The leaf buckets already store points in a contiguous array, and we group them into blocks of fixed width $W$.
We represent 
each block as a matrix $M \in \mathbb{R}^{d\times W}$ in row-major order.
By choosing $W = 8$, the matrix rows fit exactly into one 256-bit SIMD register using 32-bit floating-point arithmetic.

\begin{figure}
	\centering
	\includegraphics{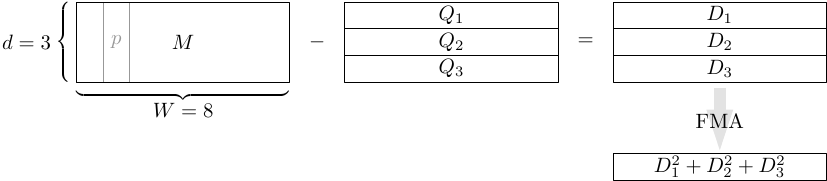}
	\caption{
		Illustration of the squared distance computation for $W=8$ points in $d=3$ dimensions, requiring $2d$ SIMD instructions.
	}\label{fig:SIMD}
\end{figure}

Let $q$ be the query point. 
To compute the squared distance to all $W$ points simultaneously, we broadcast its $i$-th coordinate $q_i$ into a SIMD register $Q_i$, creating $W$ copies of that value. 
Second, we calculate the coordinate-wise differences $D_i = Q_i - M_i$ using one SIMD subtraction per dimension, where $M_i$ is the $i$-th row of the matrix. 
We then square and sum these differences element-wise into an accumulator vector $\mathrm{acc}$ using one fused multiply-add (FMA) instruction per dimension, effectively computing $\mathrm{acc} = \sum_{i=1}^d D_i^2$ element-wise. 
This requires $2d$ SIMD instructions ($d$ subtractions and $d$ FMAs) to evaluate the distance to $W$ vectors; see~\cref{fig:SIMD}.

For higher dimensions, we reduce the instruction count further by rewriting the squared distance $\|p - q\|^2$ 
between a candidate point $p$ and the query point $q$ as $(p^\top p + q^\top q) - 2p^\top q$ and precomputing the squared norms $p^\top p$, 
as suggested by Chen and Güttel~\cite{chenFastExactFixedradius2024}.
By storing the halved squared norms, we also avoid a multiplication by $2$ during the accumulation phase:
\[
	\|p - q\|^2 = (p^\top p + q^\top q) - 2p^\top q \le r^2 \Longleftrightarrow \left(\frac{p^\top p}{2} + \frac{q^\top q}{2}\right) - p^\top q \le \frac{r^2}{2}.
\]
Computing the left side of the inequality for all $W$ points simultaneously requires only $d$ FMA instructions.
However, it requires loading the precomputed values $\frac{p^\top p}{2}$ from memory.
Due to this overhead, it only becomes faster in practice for $d \geq 6$.

To break data dependencies and maximize instruction throughput, we split the computation across \num{2} independent accumulators. 
We initialize them to the SIMD vectors containing $\frac{p^\top p}{2}$ and $\frac{q^\top q}{2}$ respectively, and alternate the subtraction of the coordinate-wise products of $p^\top q$ using the previously allocated SIMD registers $M_i$ and $Q_i$. 
For $d > 30$, utilizing \num{4} accumulation registers provides additional performance gains by further hiding instruction latency; see the detailed analysis in \cref{sec:analys-perf-gains}.
Note that the dot-product reformulation potentially introduces numerical instabilities.
We therefore add a small fixed $\varepsilon=10^{-4}$ to the query radius in our implementation.
Across all our test cases, this returned every point found by a naive reference implementation. 

After computing all candidate distances, the indices corresponding to points fulfilling $\|p - q\| \le r$ must be compacted into the result buffer.
On CPUs with AVX-512F~\cite{Intela,Intel} support, we use the \texttt{vcompressps}~\cite{Intelb} instruction to gather the indices in a single step.


\subparagraph{Memory Layout.}
In addition to vectorized arithmetic, we optimize for cache-efficiency.
Inner nodes are extremely lightweight, storing only the 32-bit floating-point split value of the dividing hyperplane.
In contrast, leaf nodes are heavy, containing the SNN bucket with metadata and point range indices.
Therefore, we keep leaves and inner nodes in separate arrays, and use for the inner nodes an implicit tree layout analogous to binary heaps.

%

Beyond the memory layout, we also optimize memory access patterns. 
During a query, a traditional KD-tree implementation frequently switches between traversing the tree and processing distance computations in the leaf nodes. 
This causes cache misses, as the CPU evicts the tree structure from the cache to load point coordinates. 
To resolve this, we decouple the query into a two-phase algorithm. 
The first phase traverses the tree in depth-first search order to compute the candidate ranges for the leaves.
It identifies all ranges that intersect the query ball and collects them into an intermediate buffer.
The second phase then computes the actual distances.
Because the coordinate data is stored contiguously, iterating over the collected ranges results in a mostly linear access pattern (with skips between ranges), thereby improving the memory prefetcher performance.\footnote{
    As our access pattern is already very efficient, more complex inner node layouts yielded no further benefits in our experiments.
    For example, we observed no improvement with a van Emde Boas layout~\cite{frigoCacheObliviousAlgorithms1999, prokop1999cache}.
}


\section{Experiments}\label{sec:experiments}

In this section, we present our experiments comparing the SPRK-tree against state-of-the-art spatial data structures.
\cref{sec:experiment-setup} outlines the experimental setup, introduces the baseline data structures, and details our new graph embedding benchmark.
Following this, we compare query performance across a diverse set of benchmarks in \cref{sec:perf-comp},
showing that the SPRK-tree consistently achieves the fastest query times across most evaluated datasets.
Finally, we identify which optimizations contribute most to its efficiency in \cref{sec:analys-perf-gains}.

\subsection{Experiment Setup}
\label{sec:experiment-setup}

All benchmarks are performed single-threaded on a dual-socket Intel Xeon Gold 6144 CPU @ \qty{3.50}{\GHz} with \num{8} cores  per socket, \qty{32}{\kibi\byte} L1 and \qty{1}{\mebi\byte} L2 cache per core, and \qty{24.75}{\mebi\byte} 
L3 cache per socket, supporting AVX-512.
The system has \qty{192}{\giga\byte} 
DDR4 RAM and runs Ubuntu 24.04 LTS.
Rust code was compiled with \texttt{rustc} version 1.94.0 in release mode with \texttt{RUSTFLAGS = -Ctarget-cpu=native}.
\CC{} code was compiled with \texttt{gcc} version 14.2 using \texttt{-O3} and \texttt{-march=native}.
For the Python libraries, CPython v3.12.3 was used.



After a 3-second warm-up, each benchmark runs for at least 20 seconds and at least 10 repetitions.
We additionally verified that the measured times are consistent with hardware performance counters (\texttt{cpu\_cycles}, \texttt{cpu\_ref\_cycles}, and \texttt{instructions}).



\subparagraph{Evaluated Data Structures.}


We compare the SPRK-tree against a broad selection of state-of-the-art spatial index implementations. 
The evaluated implementations span the scikit-learn Balltree~\cite{sklearn}, a VP-tree implemented in Rust~\cite{vp-tree}, the Boost R-tree~\cite{boost}, and SNN~\cite{chenFastExactFixedradius2024}.
Since KD-trees proved to be the strongest category overall, we include KD-tree implementations across multiple languages: the Rust crates Kiddo~\cite{donnellysdd2026}, nabo~\cite{elsebergcomparison} and Neighbourhood~\cite{hominusneighbourhood2025}, the \CC{} libraries nanoflann~\cite{nanoflann} and CGAL~\cite{CGAL}, and the scikit-learn KD-tree~\cite{sklearn}.

In addition to our SPRK-tree, we provide a brute-force baseline and our own Orthtree and uniform grid implementations.
Notably, our grid implementation requires the query radius to be specified at construction time, which may be prohibitive for some applications.
For datasets with varying query radius, we build the grid using the average query radius.
To exclude performance differences resulting from the choice of implementation language, we also provide our own Rust reimplementation of SNN, evaluated in~\cref{sec:snn_comparison}.

\subparagraph{Graph Embedding Benchmark.}

As introduced in \cref{sec:intro}, we provide a new benchmark set based on graph embeddings.
We use geometric inhomogeneous random graphs (GIRGs)~\cite{bringmannGeometric2019} generated using the method in~\cite{blasiusEfficientlyGeneratingGeometric2022}.
GIRGs are a widely used model for real-world networks with heterogeneous degree distributions and community structure.
For generation, we set the average degree to \num{15}, the power-law exponent to \num{2.5}, the dimension to \num{4}, and the temperature to \num{0} (corresponding to $\alpha = \infty$).
We vary the number of vertices $n$ such that $n\in\{10^4, 10^{4.5}, 10^5, 10^{5.5}, 10^6 \}$ and use \num{3} different seeds for each parameter configuration.

To obtain a data set from these graphs, we compute embeddings using the WEmbed algorithm~\cite{blasiusWeighted2025}.
The algorithm iteratively adjusts vertex positions starting from a random initial state.
In each iteration, the algorithm performs $n$ spherical range queries, finding the set of vertices that are within a certain distance for each vertex.
The query radius depends on the vertex degree, with high-degree vertices having a larger query radius.
To obtain data sets with varying dimension, we set the embedding dimension $d$ from \numrange{2}{16} and additionally include \num{32}.
This results in $5 \cdot 3 \cdot 16 = 240$ different embeddings across all graph sizes.

The embedding algorithm updates the vertex positions for $t$ iterations until convergence (averaging $t\approx 530$), yielding $t$ different position and query sets per embedding.
For each run, we choose the positions from the latest iteration divisible by \num{100} to include in our benchmark,
as later iterations feature less random and more interesting vertex positions and are usually more expensive.
To normalize the number of queries,
we choose at most \SI{10}{k} queries per embedding, sorting the vertices by degree and then choosing every $k$-th vertex for $k = n / 10^4$.

We also tested other parameter configurations during generation, but observed similar performance trends across all configurations; see \cref{sec:embedding_parameters}.
For readability, we only show results for $n\in\{10^4, 10^5, 10^6\}$ in our evaluation. For benchmarks spanning all vertex counts see \cref{sec:embedding_additional}.

\subsection{Performance Comparison}
\label{sec:perf-comp}

In this section, we compare the query performance of the evaluated data structures.
We begin with the graph embedding benchmark, which motivated the design of the SPRK-tree.
To evaluate the SPRK-tree's applicability beyond graph embedding and obtain a more robust comparison, we consider four additional categories: uniform point distributions, high-dimensional real-world data, clustering workloads, and Point of Interest (POI) queries.
For these, we also provide unaggregated results in \cref{sec:extern-unaggregated}.

\subparagraph{Graph Embedding.}

\begin{figure}
	\centering
	\includegraphics{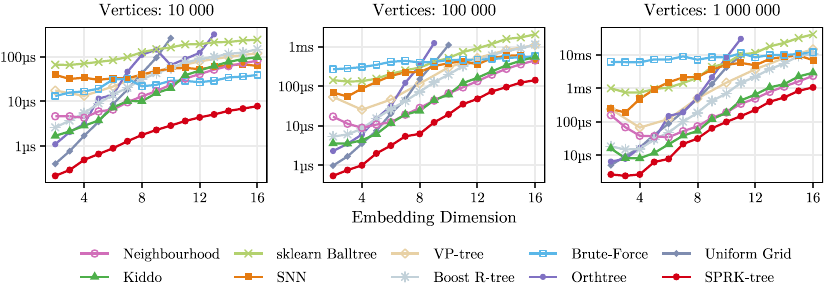}
	\caption{
	        Average time per query for our graph embedding benchmark, with varying number of vertices and embedding dimensions. The plots show a subset of structures for legibility.
	}\label{fig:benchmark_fixed_n}
\end{figure}

\begin{table}[tb]
	\centering
	\caption{
		Query time in \unit{\us} for a single spherical range query in an embedded graph for varying number of vertices and embedding dimensions.
	 	The \textbf{fastest} and \underline{second-fastest} performing methods for each combination are highlighted. The first section contains KD-tree implementations, the second section contains other methods and the third section contains our own implementations.
	 }\label{table:bench-table}
	\begin{tabular}{lrrrrrrrrr}
\toprule
num vertices & \multicolumn{3}{c}{\num{10000}} & \multicolumn{3}{c}{\num{100000}} & \multicolumn{3}{c}{\num{1000000}} \\
dimensions & \multicolumn{1}{c}{\num{2}} & \multicolumn{1}{c}{\num{8}} & \multicolumn{1}{c}{\num{16}} & \multicolumn{1}{c}{\num{2}} & \multicolumn{1}{c}{\num{8}} & \multicolumn{1}{c}{\num{16}} & \multicolumn{1}{c}{\num{2}} & \multicolumn{1}{c}{\num{8}} & \multicolumn{1}{c}{\num{16}} \\
\midrule
$\text{sklearn KD-tree}$ & \num{67.7} & \num{110.4} & \num{316.8} & \num{144.9} & \num{265.9} & \num{2038.0} & \num{1009.5} & \num{1171.7} & \num{14708.7} \\
$\text{CGAL KD-tree}$ & \num{4.3} & \num{50.5} & \num{253.5} & \num{8.6} & \num{93.8} & \num{1248.1} & \num{29.1} & \num{153.7} & \num{7734.9} \\
$\text{nanoflann}$ & \num{8.6} & \num{15.7} & \num{121.2} & \num{27.5} & \num{41.6} & \num{869.4} & \num{183.6} & \num{189.0} & \num{8439.3} \\
$\text{nabo}$ & \num{5.6} & \num{18.9} & \num{135.3} & \num{15.0} & \num{34.9} & \num{668.7} & \num{50.4} & \num{65.6} & \num{2866.3} \\
$\text{Neighbourhood}$ & \num{4.6} & \num{13.0} & \num{81.3} & \num{17.0} & \num{28.1} & \underline{\num[detect-all]{448.7}} & \num{156.9} & \num{72.3} & \underline{\num[detect-all]{2339.3}} \\
$\text{Kiddo}$ & \num{1.7} & \underline{\num[detect-all]{10.1}} & \num{98.0} & \num{3.6} & \underline{\num[detect-all]{23.8}} & \num{549.6} & \num{15.7} & \underline{\num[detect-all]{52.5}} & \num{2900.8} \\
\midrule
$\text{sklearn Balltree}$ & \num{66.1} & \num{126.0} & \num{239.0} & \num{142.1} & \num{294.9} & \num{2041.0} & \num{990.5} & \num{2036.5} & \num{40821.5} \\
$\text{SNN}$ & \num{39.2} & \num{39.6} & \num{62.2} & \num{69.2} & \num{227.3} & \num{451.5} & \num{238.8} & \num{2174.0} & \num{6875.8} \\
$\text{VP-tree}$ & \num{17.5} & \num{38.0} & \num{119.3} & \num{53.5} & \num{116.0} & \num{1096.5} & \num{236.0} & \num{493.9} & \num{14639.7} \\
$\text{Boost R-tree}$ & \num{2.6} & \num{30.9} & \num{146.6} & \num{5.3} & \num{71.7} & \num{1129.7} & \num{18.1} & \num{183.8} & \num{10543.4} \\
\midrule
$\text{Brute-Force}$ & \num{13.2} & \num{21.8} & \underline{\num[detect-all]{39.0}} & \num{268.0} & \num{398.8} & \num{570.8} & \num{6133.0} & \num{7111.6} & \num{11808.4} \\
$\text{Orthtree}$ & \num{0.6} & \num{55.1} & \num{1175.6} & \num{1.2} & \num{165.7} & \num{2162.7} & \num{3.2} & \num{269.9} & oom \\
$\text{Uniform Grid}$ & \textbf{\num[detect-all]{0.2}} & \num{21.3} & oom & \textbf{\num[detect-all]{0.5}} & \num{76.0} & oom & \textbf{\num[detect-all]{2.5}} & \num{239.1} & oom \\
$\text{SPRK-tree}$ & \textbf{\num[detect-all]{0.2}} & \textbf{\num[detect-all]{1.8}} & \textbf{\num[detect-all]{7.8}} & \textbf{\num[detect-all]{0.5}} & \textbf{\num[detect-all]{6.2}} & \textbf{\num[detect-all]{141.7}} & \underline{\num[detect-all]{2.6}} & \textbf{\num[detect-all]{31.1}} & \textbf{\num[detect-all]{1056.5}} \\
\bottomrule
\end{tabular}
\end{table}

\cref{fig:benchmark_fixed_n} illustrates query times across embedding dimensions and vertex counts for a subset of data structures,
including the best performing implementation for each algorithm.
The SPRK-tree achieves the best query time in every tested configuration, across all vertex counts and dimensions.
For $n=\SI{10}{k}$ and $d \ge 12$, all structures except the SPRK-tree are outperformed by brute-force, an expected consequence of the curse of dimensionality.
Here, the SPRK-tree likely prevails due to better-optimized constant factors.
SNN generally exhibits similar performance to brute-force for high dimensions.
We observe that SNN performs worse in dimensions not divisible by $4$,
which is likely due to the 
BLAS~\cite{OpenMathLib2026} implementation using a $4\times4$ kernel for the \texttt{sgemv} subroutine.
Furthermore, at \SI{1}{M} vertices several structures (Kiddo, Boost R-tree, Neighbourhood, and scikit-learn Balltree) are slower in 2 dimensions than in 4. 
This might be an artifact of the embedding benchmark,
where lower dimensions often have a high density of points just outside the query radius,
thereby slowing structures that are unable to efficiently prune these points.

\cref{table:bench-table} lists the comprehensive results for all evaluated structures.
Most competitors not shown in \cref{fig:benchmark_fixed_n} are clearly outperformed by the remaining structures.
In higher dimensions the grid and Orthtree ran out of memory (oom). 
For construction times see \cref{sec:index_construction}.

\subparagraph{Uniform Distributions.}

\begin{figure}
\centering
  \includegraphics{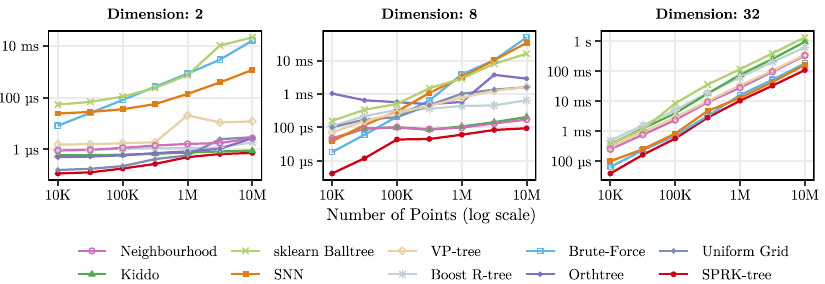}
  \caption{Query time for uniformly distributed points in a unit hypercube. The radius is chosen such that the expected number of neighbors is \num{15}.}\label{fig:distributions}
\end{figure}

Next, we evaluate a benchmark of uniformly distributed points within a unit hypercube. 
We vary the number of points from \SI{10}{k} to \SI{10}{M} and the dimension from \num{2} to \num{32}.
For each point set, we perform \SI{10}{k} queries with the query radius set such that the expected number of returned points is \num{15}.
\cref{fig:distributions} shows the results, restricted to the best-performing implementations for readability.
The SPRK-tree is the fastest structure in every tested configuration.
Its advantage is most pronounced at low point counts and low dimensions, where it outperforms the next-best competitor by up to an order of magnitude.
This gap narrows at high dimensions with many points, maintaining a factor of roughly $\num{1.6}\times$ over brute-force at dimension \num{32} with \SI{10}{M} points.
Most tree-based methods scale similarly to the SPRK-tree, but for low point counts the SPRK-tree is significantly faster, 
likely due to better cache performance; see \cref{sec:uniform-low-level-optimizations} for additional details.

The relative ranking of the competitors shifts with dimensionality.
In low dimensions (\num{2} and \num{8}), Kiddo and the Boost R-Tree are close to the SPRK-tree, while SNN is significantly slower.
For larger dimensions, 
the more sophisticated tree-based structures tend to be increasingly outperformed by brute-force,
and SNN converges toward brute-force performance.

\subparagraph{High-Dimensional Real-World Datasets.}

\begin{table}[tb]
	\centering
	\caption{Real-world query benchmarks. 
  Running time is in \unit{\ms} (nearest neighbor) or \unit{\ns} (clustering, POI) averaged over all queries and radii. 
  The \textbf{fastest} and \underline{second-fastest} methods are highlighted. 
  For unaggregated results, see \cref{sec:extern-unaggregated}.
  In some cases, the Orthtree runs out of memory (oom).
}\label{tab:extern}
	\begin{tabular}{l@{\,$\,$}rr@{\,$\,$}r@{\,$\,$}r@{\,$\,$}r@{\,$\,$}r@{\,$\,$}r@{\,$\,$}r@{\,$\,$}r@{\,$\,$}}
\toprule
Data Set & \multicolumn{1}{c}{$n$} & \multicolumn{1}{c}{$d$} & \multicolumn{1}{r}{\text{Brute-Force}} & \multicolumn{1}{r}{\text{Orthtree}} & \multicolumn{1}{r}{\text{Kiddo}} & \multicolumn{1}{r}{\text{Balltree}} & \multicolumn{1}{r}{\text{SNN}} & \multicolumn{1}{r}{\text{SPRK}} \\
\midrule
\multicolumn{7}{l}{\textbf{Real-World high-dimensional datasets (in \qty{}{\milli\second}) \cite{aumullerANNBenchmarks2020}}} \\
\midrule
SIFT1M  & \SI{100}{k} & \num{128} & \num{5.8} & oom & \num{15.3} & \num{19.9} & \textbf{\num[detect-all]{2.1}} & \underline{\num[detect-all]{2.3}} \\
SIFT10K  & \SI{25}{k} & \num{128} & \num{0.9} & oom & \num{2.2} & \num{4.1} & \textbf{\num[detect-all]{0.3}} & \underline{\num[detect-all]{0.4}} \\
GIST  & \SI{1}{M} & \num{960} & \num{433.8} & oom & \num{2591.8} & \num{1415.4} & \textbf{\num[detect-all]{183.2}} & \underline{\num[detect-all]{255.4}} \\
GloVe100  & \SI{1}{M} & \num{100} & \num{55.4} & oom & \num{49.5} & \num{295.7} & \underline{\num[detect-all]{29.7}} & \textbf{\num[detect-all]{4.4}} \\
Deep1B  & \SI{10}{M} & \num{96} & \textbf{\num[detect-all]{308.0}} & oom & \num{2831.2} & \num{2700.9} & \num{453.7} & \underline{\num[detect-all]{312.2}} \\
F-MNIST  & \SI{60}{k} & \num{784} & \num{21.2} & oom & \num{41.0} & \num{60.8} & \textbf{\num[detect-all]{5.9}} & \underline{\num[detect-all]{6.0}} \\
\midrule
\multicolumn{7}{l}{\textbf{Clustering (in \qty{}{\nano\second}) \cite{duaUCI2017}}} \\
\midrule
Banknote  & \SI{1.5}{k} & \num{4} & \num{1750.6} & \underline{\num[detect-all]{294.8}} & \num{672.4} & \num{57833.6} & \num{21593.0} & \textbf{\num[detect-all]{99.8}} \\
Wine  & \SI{0.2}{k} & \num{13} & \underline{\num[detect-all]{1131.4}} & \num{204595.4} & \num{1922.6} & \num{54822.0} & \num{18907.0} & \textbf{\num[detect-all]{143.0}} \\
Dermatology  & \SI{0.4}{k} & \num{34} & \underline{\num[detect-all]{3023.2}} & oom & \num{10815.4} & \num{65718.8} & \num{23220.4} & \textbf{\num[detect-all]{1136.4}} \\
Ecoli  & \SI{0.3}{k} & \num{7} & \num{884.4} & \num{1018.4} & \underline{\num[detect-all]{816.6}} & \num{54875.8} & \num{18502.8} & \textbf{\num[detect-all]{78.6}} \\
\midrule
\multicolumn{7}{l}{\textbf{Point of Interest Search (in \qty{}{\nano\second}) \cite{OsmGermany}}} \\
\midrule
ATM  & \SI{12}{k} & \num{2} & \num{10189.0} & \underline{\num[detect-all]{242.0}} & \num{269.0} & \num{53160.2} & \num{20069.5} & \textbf{\num[detect-all]{86.0}} \\
Bakery  & \SI{33}{k} & \num{2} & \num{27424.0} & \underline{\num[detect-all]{454.2}} & \num{648.8} & \num{69502.0} & \num{23148.8} & \textbf{\num[detect-all]{112.0}} \\
Parking  & \SI{769}{k} & \num{2} & \num{637569.5} & \underline{\num[detect-all]{1990.5}} & \num{3199.8} & \num{605426.2} & \num{66871.8} & \textbf{\num[detect-all]{619.5}} \\
Bus Stop  & \SI{761}{k} & \num{2} & \num{628722.5} & \underline{\num[detect-all]{1481.2}} & \num{2094.8} & \num{512312.8} & \num{51150.8} & \textbf{\num[detect-all]{444.8}} \\
Restaurant  & \SI{103}{k} & \num{2} & \num{84881.0} & \underline{\num[detect-all]{598.0}} & \num{745.0} & \num{109636.5} & \num{27474.8} & \textbf{\num[detect-all]{167.5}} \\
\bottomrule
\end{tabular}
\end{table}

We further evaluate data structures on high-dimensional real-world datasets from the benchmark collection of Aumüller, Bernhardsson and Faithfull~\cite{aumullerANNBenchmarks2020}.
This collection features datasets with up to \SI{10}{M} points and up to \num{960} dimensions; see \cref{tab:extern} (first section).
Some datasets exhibit very high dimension-to-point ratios; the small SIFT dataset, for instance, has \num{128} dimensions but only \SI{25}{k} points.

Based on our previous benchmark results, one would expect tree-based methods to struggle heavily in this regime. 
While other tree-based implementations indeed fall behind by an order of magnitude or more, 
the SPRK-tree remains competitive, converging toward brute-force and SNN performance levels.
SNN is the fastest method on four of the six nearest neighbor datasets, while the SPRK-tree consistently places second.
We attribute the SPRK-tree's strong performance 
mostly to the initial SVD rotation and optimized distance computation. 
The SVD is also applied by SNN, explaining the similar performance.

\subparagraph{Clustering.}

We also evaluate the clustering benchmarks introduced in the SNN paper~\cite{chenFastExactFixedradius2024}.
These benchmarks use DBSCAN~\cite{esterDensityBased}, a density-based clustering algorithm whose core operation is a fixed-radius range query for every point in the dataset.
We isolate the all-to-all query component to measure spatial index performance independently of the clustering logic
(note that~\cite{chenFastExactFixedradius2024} reports the end-to-end runtime).
The datasets are small, containing \numrange{200}{4500} points, but span dimensions from \numrange{4}{256}.

As shown in the second section of \cref{tab:extern}, the SPRK-tree achieves the best query time on all clustering datasets, often by a wide margin.
Brute-force and Kiddo alternate as the second-best method depending on the dimensionality: Kiddo places second on the lower-dimensional datasets (Banknote, Ecoli), while brute-force takes second on the higher-dimensional ones (Dermatology, Wine).
SNN performs poorly across all clustering datasets, likely due to constant overheads that dominate at these small point counts.

\subparagraph{POI Queries.}
To evaluate performance on geographic data, we benchmark fixed-radius POI queries, 
e.g., finding all bus stops within \qty{500}{\meter} of a given location.
These datasets are two-dimensional but exhibit highly heterogeneous density, with dense clusters in cities and sparse coverage in rural areas.
We use data from the OpenStreetMap project for Germany~\cite{OsmGermany},
with five categories:
ATMs ($\approx \SI{12}{k}$), bakeries ($\approx \SI{33}{k}$), parking facilities ($\approx \SI{769}{k}$), bus stops ($\approx \SI{761}{k}$), and restaurants ($\approx \SI{103}{k}$).
Each query set is constructed by using POIs of one category as dataset and POIs of a different category as query points.
Queries are run at radii of \qty{500}{\meter}, \qty{1000}{\meter}, \qty{2000}{\meter}, and \qty{5000}{\meter}, and times are averaged across all radii and queries.

As shown in the POI section of \cref{tab:extern}, the SPRK-tree achieves the best query time across all five datasets.
SNN is \num{2} to \num{3} orders of magnitude slower in this setting.
The reason is likely that in \num{2} dimensions, the pruning of tree-based methods is more effective than projecting onto a single axis.
Among tree-based methods, the Orthtree is the strongest competitor, yet the SPRK-tree outperforms it by a factor of $2.8 \times$ on the smallest dataset (ATM, \SI{12}{k} points) and up to $3.3 \times$ on larger datasets (Bus Stop, \SI{761}{k} points).
We attribute this to spherical pruning, which reduces the distance checks by roughly \SI{50}{\%} in \num{2} dimensions.


\subsection{Breakdown of Performance Gains}
\label{sec:analys-perf-gains}

In this section, we analyze the individual contributions of the different optimizations (both algorithmic and low-level) to the overall performance of the SPRK-tree.
We first measure the reduction in distance checks achieved by radius reduction and SNN buckets,
followed by evaluating the SIMD and memory layout optimizations.

\subparagraph{Algorithmic Optimizations.}

\begin{figure}
	\centering
	\includegraphics{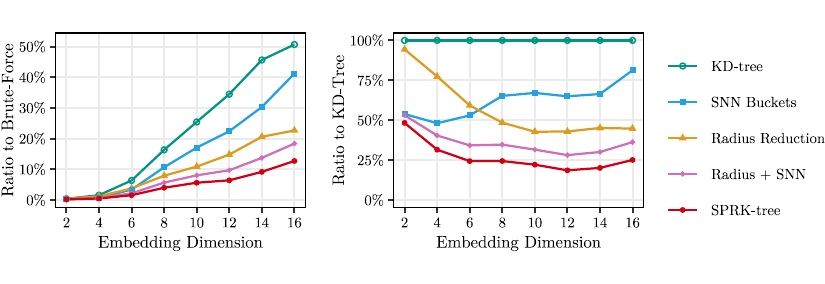}
    \vspace{-15pt}
	\caption{Effect of radius reduction and SNN buckets for a \SI{100}{k} point dataset.
	The y-axis shows the ratio of distance checks performed compared to brute-force (left) and a KD-tree (right).
}\label{fig:distance_checks_atree}
\end{figure}

Recall that the SPRK-tree is a KD-tree with radius reduction and SNN in the leaves (Section~\ref{sec:SPRK-tree}).
In \cref{fig:distance_checks_atree}, we present an ablation study for these optimizations on the \SI{100}{k} embedding dataset.
We measure the number of distance computations performed by the data structure, which is independent of low-level optimizations.

The baseline is our KD-tree implementation without radius reduction or SNN (\texttt{KD-tree}) as described in \cref{sec:kd-tree}.
Using only radius reduction (\texttt{Radius Reduction}) cuts the number of distance checks significantly, achieving up to \SI{50}{\%} fewer checks in higher dimensions.
On the other hand, SNN buckets in isolation (\texttt{SNN Buckets}) are most effective in lower dimensions, where they achieve a similar reduction of \SI{50}{\%}.
The full benefit is realized by combining both techniques (\texttt{Radius + SNN}) and also using the reduced radius to further narrow the SNN candidate range (representing the full \texttt{SPRK-Tree}).
This brings the total number of distance checks down to about \SI{25}{\%} of a regular KD-tree for $d \ge 6$.

\subparagraph{Distance Computation.}

\begin{figure}
    \centering
    \includegraphics{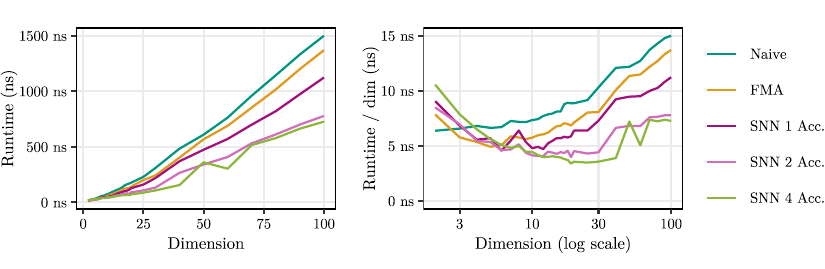}
    \vspace{-15pt}
    \caption{
        Effect of fused multiply-add operations and different numbers of accumulation registers.
        Runtime is measured for computing 128 distances to a single query point.
        Right: performance normalized by dimension with a logarithmic x-axis.
    }\label{fig:acc_microbench}
\end{figure}

To evaluate the low-level optimizations of the distance computation introduced in \cref{sec:perf_engineering}, we conduct a microbenchmark.
We measure the computation of 128 distances to a single query point across varying dimensions. 
We compare the direct distance computation without optimizations (\texttt{naive}), the approach using fused multiply-add instructions (\texttt{FMA}),
the inner product reformulation (\texttt{SNN Acc}), and the variants using \num{2} and \num{4} accumulation registers (\texttt{SNN 2 Acc} and \texttt{SNN 4 Acc}).
The results are shown in \cref{fig:acc_microbench}.

The fused multiply-add approach provides moderate improvement over the naive baseline, reducing runtime by roughly \SI{10}{\%} on average.
The inner product reformulation initially performs worse, but becomes faster than \texttt{FMA} at \num{6} dimensions.
In low dimensions, loading the precomputed norms from memory likely outweighs the benefits of reduced arithmetic operations.
The variants with multiple accumulators provide additional improvements in higher dimensions.
For the dimensions targeted by our main benchmarks ($d \leq 16$), \num{2} accumulators offer a good trade-off, with a speedup of up to \SI{35}{\%} over one accumulator.
This is likely due to fewer data dependencies, allowing the CPU to better hide instruction latency.

Based on these insights, the SPRK-tree uses a dimension-dependent case distinction.
For $d < 6$ the SPRK-tree uses the FMA-based distance computation.
For dimensions up to $d = 32$, it switches to the inner product method with \num{2} accumulators.
Beyond that, it uses \num{4} accumulators.
We note that the optimal number of accumulators is architecture-dependent, as it interacts with register pressure and the compiler's auto-vectorization logic.

\subparagraph{Cumulative Optimization Effects.}

\begin{figure}
    \centering
    \includegraphics{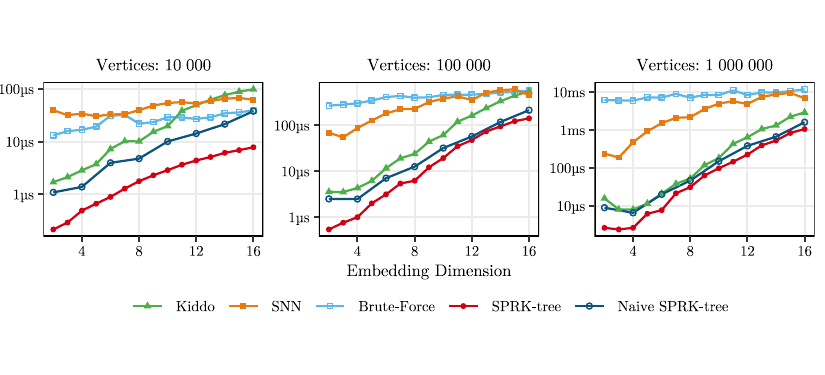}
    \caption{SPRK-tree query time with and without the optimizations from \cref{sec:perf_engineering}. The naive variant retains all algorithmic optimizations but disables low-level performance optimizations.}\label{fig:atree_without_opt}
\end{figure}

In \cref{fig:atree_without_opt}, we compare the full SPRK-tree against a naive variant that retains all algorithmic optimizations but disables the low-level optimizations from \cref{sec:perf_engineering},
including the memory layout and access pattern optimizations.
The gains are most pronounced for low point counts, where for example even writing results to the output buffer constitutes a significant fraction of the total query time.
Notably, the naive SPRK-tree already performs well for small point counts and high dimensions.
This demonstrates the effectiveness of the algorithmic optimizations alone,
with the low-level engineering further improving the performance by up to an order of magnitude at small point counts.

\section{Conclusion}\label{sec:conclusion}

While developed primarily for repelling-force computation in graph embeddings, the SPRK-tree proves to be an efficient, general-purpose spatial index across a wide range of applications.
Even in extreme scenarios, such as high-dimensional real-world datasets, the SPRK-tree remains competitive with the best-performing methods.
This is achieved with a combination of algorithmic optimizations, such as the spherical pruning and SNN buckets, and low-level engineering that maximizes hardware utilization.
Integrating the SPRK-tree should allow to substantially improve the performance of applications across various domains.

Another common strategy for high-dimensional regimes is to resort to approximate queries.
Our initial attempts in this direction were rather unpromising, due to a significant drop in embedding quality for only small performance gain.
However, additional work is warranted to fully evaluate the potential of this direction.

\bibliography{references.bib}

\clearpage
\appendix

\section{SPRK-Tree Query Procedure}\label{sec:sprk_query}

In \cref{algo:node} and \cref{algo:leaf}, we provide the pseudocode for the SPRK-tree query procedure, which is described in \cref{sec:SPRK-tree}. Since the SPRK-tree is balanced and every branch has the same depth, whether a node is a leaf or an internal node can be determined by its depth.

\begin{algorithm}[tbhp]
	\caption{SPRK-tree procedure to query an internal node}\label{algo:node}
	\SetKwFunction{FQueryInnerNode}{QueryInternalNode}
	\SetKwFunction{FQueryNode}{QueryNode}
	\SetKwProg{Fn}{Function}{:}{}
	\Fn{\FQueryInnerNode{$v$, $q$, ${\|\Delta\|}^2$, $r$, $\Delta$}}{
		$k, s \gets$ split dimension and split value of $v$\;
		$v_1, v_2 \gets$ child of $v$ on same side as $q$ and opposite side as $q$\;
		\FQueryNode{$v_1$, $q$, ${\|\Delta\|}^2$, $r$, $\Delta$}
		\tcp*{Query subtree on the same side as $q$}
		$\delta \gets s - q_k$
		\tcp*{Update reduced radius for opposite subtree}
		${\|\Delta\|}^2 \gets {\|\Delta\|}^2 - {\Delta[k]}^2 + \delta^2$\;
		$\Delta[k] \gets \delta$
		\tcp*{Update the $k$-th component of $\Delta$}
		\lIf{${\|\Delta\|}^2 \leq r^2$}{
		\FQueryNode{$v_2$, $q$, ${\|\Delta\|}^2$, $r$, $\Delta$}}
	}
\end{algorithm}

\begin{algorithm}[tbhp]
	\caption{SPRK-tree procedure to query a leaf node}\label{algo:leaf}
	\SetKwFunction{FQuerySNN}{QueryLeaf}
	\SetKwFunction{FQuerySNNInner}{SNN}
	\SetKwProg{Fn}{Function}{:}{}
	\Fn{\FQuerySNN{$\ell$, $q$, ${\|\Delta\|}^2$, $r$, $\Delta$}}{
		$k \gets$ split dimension of $\ell$\;
		$r_{\text{SNN}} \gets \sqrt{r^2 - {\|\Delta\|}^2  + {\Delta[k]}^2}$\;
		$\text{min}, \text{max} \gets q_k - r_{\text{SNN}},\; q_k + r_{\text{SNN}}$\;
		$T \gets \{ p\in P_\ell \mid \text{min} \leq p_k \leq \text{max} \}$ using lookup table\;
		\Return $\{ p \in T \mid \|p-q\|^2 \leq r^2 \}$\;
	}
\end{algorithm}

\section{Graph Embedding Parameters}\label{sec:embedding_parameters}
\begin{figure}[tbhp]
	\centering
	\includegraphics{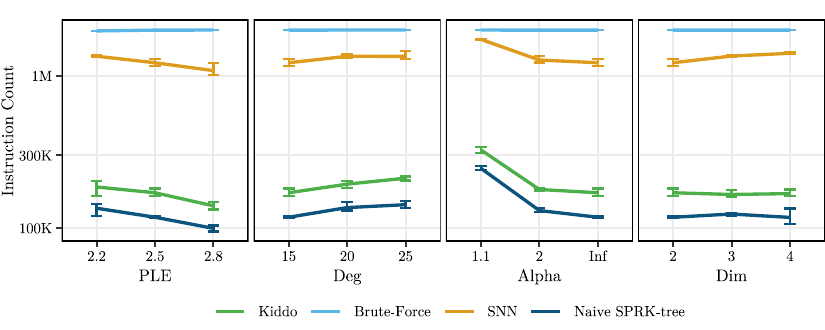}
	\caption{Effect of graph parameters on relative query time. The order of all structures remains largely consistent.}\label{fig:parameters_line}
\end{figure}

In \cref{sec:experiments}, we described our new graph embedding benchmark set. 
In \cref{fig:parameters_line}, we provide our justification for focusing on a single parameter configuration in the main benchmark. 
We benchmarked the effect of varying the GIRG generation parameters (average degree, power-law exponent, temperature, and sampling dimension) and observed similar performance trends across all configurations. 
Most notably, the change in relative performance of the different data structures remains largely within the margin of error across all configurations.

\section{Additional Graph Embedding Results}\label{sec:embedding_additional}
\begin{figure}[h]
	\centering
	\includegraphics{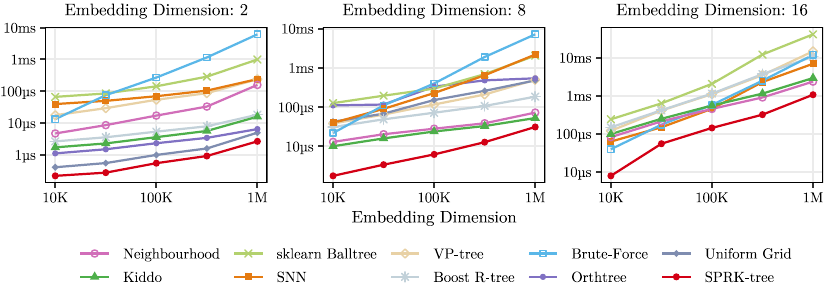}
	\caption{Average time per query for our graph embedding benchmark, with varying number of vertices and embedding dimensions.}\label{fig:benchmark_all_n}
\end{figure}

In \cref{fig:benchmark_fixed_n} we evaluated the structures on a subset of the vertex counts of our benchmarking set to highlight the interesting performance characteristics. 
In \cref{fig:benchmark_all_n} we show the results for all vertex counts for dimensions \num{2}, \num{8}, and \num{16}.
The analyzed trends remain consistent with the subset shown in \cref{fig:benchmark_fixed_n}, with the SPRK-tree achieving the best query time across all configurations.

\section{Index Construction Time}\label{sec:index_construction}
\begin{figure}[h]
	\centering
	\includegraphics{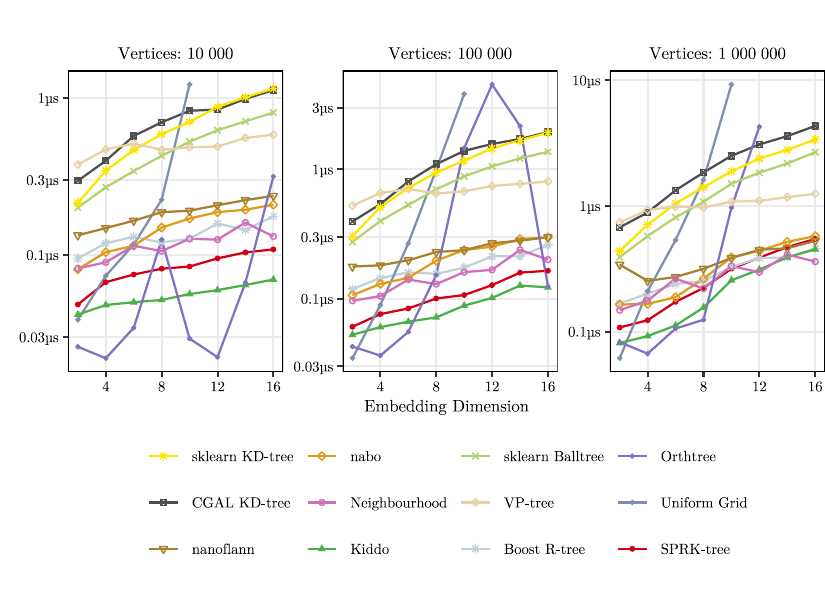}
	\caption{Index construction time with varying number of vertices and embedding dimensions.}\label{fig:index_construction}
\end{figure}

In the context of graph embedding, the index needs to be reconstructed from scratch in every iteration of the embedding algorithm, making the index construction time a relevant performance metric. 
The index construction time of the benchmarked data structures, on the other hand, is negligible compared to the query time of the $n$ queries performed in each iteration, so we do not include it in the main benchmark. 
\cref{fig:index_construction} shows the index construction time of all evaluated data structures for the embedding benchmark. The implementations were tested in their default configurations, some of which use multithreading by default.
The SPRK-tree has a competitive construction time, staying within the same order of magnitude as the other fast-constructing, tree-based methods.
The Orthtree exhibits a noteworthy construction time with a steep increase with the number of dimensions but periodic large drops. The steepness naturally correlates to the exponential increase in the number of cells with increasing dimension. 
The periodic drops on the other hand are likely due to the fact that the depth of the tree decreases.
Both the Orthtree and the uniform grid run out of memory for larger point counts at higher dimensions.

\section{Effect of Low-Level Optimizations for the uniform distribution benchmark}\label{sec:uniform-low-level-optimizations}
\begin{figure}[h]
	\centering
	\includegraphics{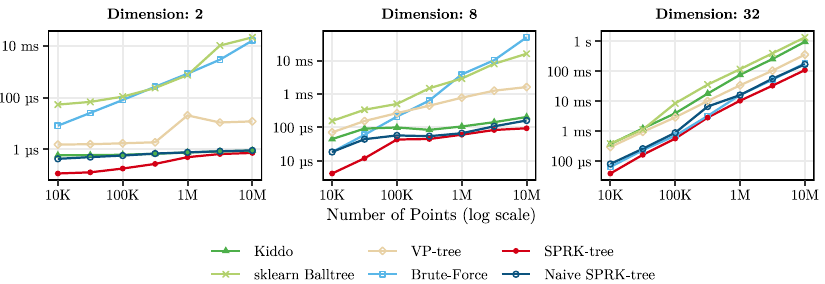}
	\caption{Query time for uniformly distributed points in a unit hypercube including the naive SPRK-tree variant without the low-level optimizations from \cref{sec:perf_engineering}. The radius is chosen such that the expected number of neighbors is \num{15}.}\label{fig:distributions_naive}
\end{figure}

In \cref{sec:experiments} we highlighted the performance gains of the SPRK-tree on the uniform distribution benchmark, especially at low point counts and low dimensions. 
We attribute this to better cache performance, which is a result of the low-level optimizations described in \cref{sec:perf_engineering}. 
In \cref{fig:distributions_naive} we demonstrate the effect of these optimizations for this benchmark by comparing the fully optimized SPRK-tree to a naive variant that retains all algorithmic optimizations but disables the low-level optimizations from \cref{sec:perf_engineering}. 
We observe that the optimizations provide a significant performance boost for lower point counts in 2D, emphasizing that the slope of the fully optimized SPRK-tree is due to better cache performance rather than worse scaling.

This can also be explained mathematically by considering the cache occupancy of the dataset.
For example, in dimension 8, each point occupies 8 times 32 bits, so \SI{10}{k}~points (\qty{312}{\kibi\byte}) fit comfortably in the L2 cache (\qty{1}{\mebi\byte}).
From \SI{10}{k} to \SI{100}{k} points the data starts to outgrow L2, causing a steep initial rise, exceeding its capacity at \SI{100}{k} points (\qty{3.1}{\mebi\byte}).
Beyond this range the scaling flattens again: although the full dataset surpasses the \qty{24.75}{\mebi\byte} L3 cache, the tree-traversal metadata remains substantially smaller and stays resident in L3.
Because cache occupancy also depends on the dimension, this transition shifts to lower point counts as dimension increases, producing steeper initial slopes at higher dimensions.


\section{Unaggregated Results for the Real-World Benchmarks.}\label{sec:extern-unaggregated}
In \cref{tab:extern} we reported the average query time across all queries and multiple radii for the high-dimensional real-world datasets, clustering, and POI benchmarks. In \cref{tab:nn_extern}, \cref{tab:clustering_extern}, and \cref{tab:poi_extern} we provide the unaggregated results.
In contrast to the aggregated results in \cref{tab:extern}, \cref{tab:poi_extern} includes the uniform grid instead of the ball tree since it provides additional interesting insights. Keep in mind that the uniform grid profits strongly from the fact that the radius is fixed within each benchmark, providing a somewhat optimistic lower bound on its performance in realistic scenarios. Nonetheless, it is noteworthy that the uniform grid is still beaten by the SPRK-tree in almost all POI benchmark configurations, only surpassing it for small radii on the smaller datasets.

\begin{table}[h]
	\centering
	\caption{Real-world spherical range queries benchmarks in high dimensions. Runtime is given in \unit{\ms}. The \textbf{best} and \underline{second best} performing methods are highlighted.}\label{tab:nn_extern}
	\begin{tabular}{@{}l@{\,$\,$}rr@{\,$\,$}r@{\,$\,$}r@{\,$\,$}r@{\,$\,$}r@{\,$\,$}r@{\,$\,$}r@{}}
\toprule
Data Set & \multicolumn{1}{c}{$n$} & \multicolumn{1}{c}{$d$} & Radius & \multicolumn{1}{c}{\text{Brute-Force}} & \multicolumn{1}{c}{\text{Kiddo}} & \multicolumn{1}{c}{\text{Balltree}} & \multicolumn{1}{c}{\text{SNN}} & \multicolumn{1}{c}{\text{SPRK}} \\
\midrule
SIFT1M  & \SI{100}{k} & \num{128} & \num{210.0} & \num{5.54} & \num{13.34} & \num{17.17} & \underline{\num[detect-all]{2.03}} & \textbf{\num[detect-all]{1.85}} \\
SIFT1M  & \SI{100}{k} & \num{128} & \num{230.0} & \num{5.34} & \num{18.59} & \num{18.81} & \textbf{\num[detect-all]{2.01}} & \underline{\num[detect-all]{2.3}} \\
SIFT1M  & \SI{100}{k} & \num{128} & \num{250.0} & \num{5.26} & \num{18.33} & \num{20.23} & \textbf{\num[detect-all]{2.02}} & \underline{\num[detect-all]{2.28}} \\
SIFT1M  & \SI{100}{k} & \num{128} & \num{270.0} & \num{5.13} & \num{13.18} & \num{20.89} & \textbf{\num[detect-all]{2.09}} & \underline{\num[detect-all]{2.39}} \\
SIFT1M  & \SI{100}{k} & \num{128} & \num{290.0} & \num{8.05} & \num{13.12} & \num{22.31} & \textbf{\num[detect-all]{2.21}} & \underline{\num[detect-all]{2.72}} \\
SIFT10K  & \SI{25}{k} & \num{128} & \num{210.0} & \num{0.96} & \num{2.2} & \num{3.53} & \textbf{\num[detect-all]{0.31}} & \underline{\num[detect-all]{0.36}} \\
SIFT10K  & \SI{25}{k} & \num{128} & \num{230.0} & \num{0.9} & \num{2.17} & \num{3.93} & \textbf{\num[detect-all]{0.31}} & \underline{\num[detect-all]{0.39}} \\
SIFT10K  & \SI{25}{k} & \num{128} & \num{250.0} & \num{0.9} & \num{2.21} & \num{4.29} & \textbf{\num[detect-all]{0.31}} & \underline{\num[detect-all]{0.43}} \\
SIFT10K  & \SI{25}{k} & \num{128} & \num{270.0} & \num{0.89} & \num{2.25} & \num{4.37} & \textbf{\num[detect-all]{0.34}} & \underline{\num[detect-all]{0.43}} \\
SIFT10K  & \SI{25}{k} & \num{128} & \num{290.0} & \num{0.89} & \num{2.24} & \num{4.47} & \textbf{\num[detect-all]{0.33}} & \underline{\num[detect-all]{0.42}} \\
GIST  & \SI{1}{M} & \num{960} & \num{0.8} & \num{450.18} & \num{2369.59} & \num{1318.36} & \textbf{\num[detect-all]{172.8}} & \underline{\num[detect-all]{228.73}} \\
GIST  & \SI{1}{M} & \num{960} & \num{0.85} & \num{422.84} & \num{2979.02} & \num{1510.28} & \textbf{\num[detect-all]{177.49}} & \underline{\num[detect-all]{247.91}} \\
GIST  & \SI{1}{M} & \num{960} & \num{0.9} & \num{426.66} & \num{2686.43} & \num{1402.26} & \textbf{\num[detect-all]{186.33}} & \underline{\num[detect-all]{263.32}} \\
GIST  & \SI{1}{M} & \num{960} & \num{0.95} & \num{431.7} & \num{2453.25} & \num{1413.04} & \textbf{\num[detect-all]{188.76}} & \underline{\num[detect-all]{266.99}} \\
GIST  & \SI{1}{M} & \num{960} & \num{1.0} & \num{437.73} & \num{2470.93} & \num{1433.22} & \textbf{\num[detect-all]{190.58}} & \underline{\num[detect-all]{270.03}} \\
GloVe100  & \SI{1}{M} & \num{100} & \num{0.94} & \num{56.01} & \num{34.47} & \num{291.37} & \underline{\num[detect-all]{30.37}} & \textbf{\num[detect-all]{3.45}} \\
GloVe100  & \SI{1}{M} & \num{100} & \num{0.97} & \num{54.28} & \num{39.83} & \num{292.71} & \underline{\num[detect-all]{30.95}} & \textbf{\num[detect-all]{3.88}} \\
GloVe100  & \SI{1}{M} & \num{100} & \num{1.01} & \num{58.04} & \num{60.35} & \num{300.59} & \underline{\num[detect-all]{28.46}} & \textbf{\num[detect-all]{4.36}} \\
GloVe100  & \SI{1}{M} & \num{100} & \num{1.04} & \num{53.36} & \num{54.16} & \num{301.59} & \underline{\num[detect-all]{29.56}} & \textbf{\num[detect-all]{5.01}} \\
GloVe100  & \SI{1}{M} & \num{100} & \num{1.07} & \num{55.32} & \num{58.87} & \num{292.27} & \underline{\num[detect-all]{29.03}} & \textbf{\num[detect-all]{5.48}} \\
Deep1B  & \SI{10}{M} & \num{96} & \num{0.69} & \underline{\num[detect-all]{312.0}} & \num{2701.79} & \num{2731.03} & \num{351.63} & \textbf{\num[detect-all]{293.52}} \\
Deep1B  & \SI{10}{M} & \num{96} & \num{0.75} & \textbf{\num[detect-all]{307.12}} & \num{3077.48} & \num{2621.94} & \num{369.2} & \underline{\num[detect-all]{312.19}} \\
Deep1B  & \SI{10}{M} & \num{96} & \num{0.82} & \textbf{\num[detect-all]{306.93}} & \num{2463.67} & \num{2744.74} & \num{540.62} & \underline{\num[detect-all]{317.95}} \\
Deep1B  & \SI{10}{M} & \num{96} & \num{0.88} & \textbf{\num[detect-all]{306.91}} & \num{2907.12} & \num{2818.33} & \num{505.08} & \underline{\num[detect-all]{318.85}} \\
Deep1B  & \SI{10}{M} & \num{96} & \num{0.94} & \textbf{\num[detect-all]{307.07}} & \num{3005.74} & \num{2588.49} & \num{502.21} & \underline{\num[detect-all]{318.37}} \\
F-MNIST  & \SI{60}{k} & \num{784} & \num{800.0} & \num{20.78} & \num{41.16} & \num{59.67} & \underline{\num[detect-all]{4.85}} & \textbf{\num[detect-all]{4.55}} \\
F-MNIST  & \SI{60}{k} & \num{784} & \num{900.0} & \num{20.88} & \num{37.47} & \num{60.45} & \underline{\num[detect-all]{5.44}} & \textbf{\num[detect-all]{5.07}} \\
F-MNIST  & \SI{60}{k} & \num{784} & \num{1000.0} & \num{21.25} & \num{38.84} & \num{60.34} & \underline{\num[detect-all]{5.93}} & \textbf{\num[detect-all]{5.9}} \\
F-MNIST  & \SI{60}{k} & \num{784} & \num{1100.0} & \num{21.38} & \num{39.3} & \num{62.12} & \textbf{\num[detect-all]{6.53}} & \underline{\num[detect-all]{6.78}} \\
F-MNIST  & \SI{60}{k} & \num{784} & \num{1200.0} & \num{21.52} & \num{48.19} & \num{61.39} & \textbf{\num[detect-all]{7.09}} & \underline{\num[detect-all]{7.47}} \\
\bottomrule
\end{tabular}
\end{table}
\begin{table}[h]
	\centering
	\caption{Real-world clustering benchmarks. Runtime in \unit{\ns}. The \textbf{best} and \underline{second best} performing methods are highlighted.}\label{tab:clustering_extern}
	\begin{tabular}{@{}l@{\,$\,$}rr@{\,$\,$}r@{\,$\,$}r@{\,$\,$}r@{\,$\,$}r@{\,$\,$}r@{\,$\,$}r@{\,$\,$}r@{}}
\toprule
Data Set & \multicolumn{1}{c}{$n$} & \multicolumn{1}{c}{$d$} & Radius & \multicolumn{1}{c}{\text{Brute-Force}} & \multicolumn{1}{c}{\text{Orthtree}} & \multicolumn{1}{c}{\text{Kiddo}} & \multicolumn{1}{c}{\text{Balltree}} & \multicolumn{1}{c}{\text{SNN}} & \multicolumn{1}{c}{\text{SPRK}} \\
\midrule
Banknote  & \SI{1.5}{k} & \num{4} & \num{0.1} & \num{1624} & \underline{\num[detect-all]{164}} & \num{219} & \num{54767} & \num{25491} & \textbf{\num[detect-all]{57}} \\
Banknote  & \SI{1.5}{k} & \num{4} & \num{0.2} & \num{1655} & \underline{\num[detect-all]{219}} & \num{396} & \num{55098} & \num{19182} & \textbf{\num[detect-all]{75}} \\
Banknote  & \SI{1.5}{k} & \num{4} & \num{0.3} & \num{1739} & \underline{\num[detect-all]{292}} & \num{664} & \num{57099} & \num{20545} & \textbf{\num[detect-all]{96}} \\
Banknote  & \SI{1.5}{k} & \num{4} & \num{0.4} & \num{1824} & \underline{\num[detect-all]{341}} & \num{916} & \num{57155} & \num{20895} & \textbf{\num[detect-all]{137}} \\
Banknote  & \SI{1.5}{k} & \num{4} & \num{0.5} & \num{1911} & \underline{\num[detect-all]{458}} & \num{1167} & \num{65049} & \num{21852} & \textbf{\num[detect-all]{134}} \\
Wine  & \SI{0.2}{k} & \num{13} & \num{2.2} & \underline{\num[detect-all]{1099}} & \num{190648} & \num{1795} & \num{54623} & \num{18629} & \textbf{\num[detect-all]{134}} \\
Wine  & \SI{0.2}{k} & \num{13} & \num{2.3} & \underline{\num[detect-all]{1111}} & \num{199223} & \num{1848} & \num{54697} & \num{18742} & \textbf{\num[detect-all]{169}} \\
Wine  & \SI{0.2}{k} & \num{13} & \num{2.4} & \underline{\num[detect-all]{1121}} & \num{208220} & \num{1905} & \num{54926} & \num{18873} & \textbf{\num[detect-all]{135}} \\
Wine  & \SI{0.2}{k} & \num{13} & \num{2.5} & \underline{\num[detect-all]{1147}} & \num{208048} & \num{1988} & \num{55108} & \num{19032} & \textbf{\num[detect-all]{138}} \\
Wine  & \SI{0.2}{k} & \num{13} & \num{2.6} & \underline{\num[detect-all]{1179}} & \num{216838} & \num{2077} & \num{54756} & \num{19259} & \textbf{\num[detect-all]{139}} \\
Dermatology  & \SI{0.4}{k} & \num{34} & \num{5.0} & \underline{\num[detect-all]{2964}} & oom & \num{10764} & \num{64381} & \num{22710} & \textbf{\num[detect-all]{1132}} \\
Dermatology  & \SI{0.4}{k} & \num{34} & \num{5.1} & \underline{\num[detect-all]{2971}} & oom & \num{10597} & \num{66008} & \num{22897} & \textbf{\num[detect-all]{1124}} \\
Dermatology  & \SI{0.4}{k} & \num{34} & \num{5.2} & \underline{\num[detect-all]{3000}} & oom & \num{10937} & \num{65162} & \num{23085} & \textbf{\num[detect-all]{1136}} \\
Dermatology  & \SI{0.4}{k} & \num{34} & \num{5.3} & \underline{\num[detect-all]{3040}} & oom & \num{10756} & \num{66889} & \num{23731} & \textbf{\num[detect-all]{1137}} \\
Dermatology  & \SI{0.4}{k} & \num{34} & \num{5.4} & \underline{\num[detect-all]{3141}} & oom & \num{11023} & \num{66154} & \num{23679} & \textbf{\num[detect-all]{1153}} \\
Ecoli  & \SI{0.3}{k} & \num{7} & \num{0.5} & \num{847} & \num{838} & \underline{\num[detect-all]{622}} & \num{54748} & \num{18320} & \textbf{\num[detect-all]{68}} \\
Ecoli  & \SI{0.3}{k} & \num{7} & \num{0.6} & \num{864} & \num{937} & \underline{\num[detect-all]{696}} & \num{54839} & \num{18378} & \textbf{\num[detect-all]{74}} \\
Ecoli  & \SI{0.3}{k} & \num{7} & \num{0.7} & \num{871} & \num{990} & \underline{\num[detect-all]{808}} & \num{54171} & \num{18210} & \textbf{\num[detect-all]{78}} \\
Ecoli  & \SI{0.3}{k} & \num{7} & \num{0.8} & \underline{\num[detect-all]{899}} & \num{1129} & \num{908} & \num{54538} & \num{18437} & \textbf{\num[detect-all]{83}} \\
Ecoli  & \SI{0.3}{k} & \num{7} & \num{0.9} & \underline{\num[detect-all]{941}} & \num{1198} & \num{1049} & \num{56083} & \num{19169} & \textbf{\num[detect-all]{90}} \\
\bottomrule
\end{tabular}
\end{table} 
\begin{table}[h]
	\centering
	\caption{Geographic point of interest search benchmarks in 2 dimensions. Runtime in \unit{\ns}. The \textbf{best} and \underline{second best} performing methods are highlighted.}\label{tab:poi_extern}
	\begin{tabular}{@{}r@{\,$\,$}r@{\,$\,$}r@{\,$\,$}r@{\,$\,$}r@{\,$\,$}r@{\,$\,$}r@{\,$\,$}r@{\,$\,$}r@{}}
\toprule
Radius & \multicolumn{1}{c}{\text{Brute-Force}} & \multicolumn{1}{c}{\text{Orthtree}} & \multicolumn{1}{c}{\text{Kiddo}} & \multicolumn{1}{c}{\text{Balltree}} & \multicolumn{1}{c}{\text{Uniform Grid}} & \multicolumn{1}{c}{\text{SNN}} & \multicolumn{1}{c}{\text{SPRK}} \\
\midrule
\multicolumn{8}{l}{\textbf{ATM (12K) queried from supermarkets}} \\
\midrule
\num{500.0} & \num{10177} & \num{146} & \num{123} & \num{52143} & \textbf{\num[detect-all]{49}} & \num{18010} & \underline{\num[detect-all]{73}} \\
\num{1000.0} & \num{10032} & \num{177} & \num{167} & \num{51568} & \textbf{\num[detect-all]{61}} & \num{18270} & \underline{\num[detect-all]{78}} \\
\num{2000.0} & \num{10076} & \num{242} & \num{255} & \num{54878} & \underline{\num[detect-all]{89}} & \num{22356} & \textbf{\num[detect-all]{84}} \\
\num{5000.0} & \num{10471} & \num{403} & \num{531} & \num{54052} & \underline{\num[detect-all]{152}} & \num{21642} & \textbf{\num[detect-all]{109}} \\
\midrule
\multicolumn{8}{l}{\textbf{Bakery (33K) queried from universities}} \\
\midrule
\num{500.0} & \num{27058} & \num{197} & \num{212} & \num{67264} & \textbf{\num[detect-all]{82}} & \num{18052} & \underline{\num[detect-all]{84}} \\
\num{1000.0} & \num{27247} & \num{316} & \num{400} & \num{66353} & \underline{\num[detect-all]{114}} & \num{19452} & \textbf{\num[detect-all]{93}} \\
\num{2000.0} & \num{27341} & \num{487} & \num{709} & \num{68633} & \underline{\num[detect-all]{155}} & \num{22730} & \textbf{\num[detect-all]{111}} \\
\num{5000.0} & \num{28050} & \num{817} & \num{1274} & \num{75758} & \underline{\num[detect-all]{249}} & \num{32361} & \textbf{\num[detect-all]{160}} \\
\midrule
\multicolumn{8}{l}{\textbf{Parking (769K) queried from hospitals}} \\
\midrule
\num{500.0} & \num{635225} & \num{794} & \num{815} & \num{532944} & \underline{\num[detect-all]{441}} & \num{26809} & \textbf{\num[detect-all]{259}} \\
\num{1000.0} & \num{634616} & \num{1175} & \num{1364} & \num{508476} & \underline{\num[detect-all]{574}} & \num{36600} & \textbf{\num[detect-all]{367}} \\
\num{2000.0} & \num{638349} & \num{1974} & \num{3001} & \num{524187} & \underline{\num[detect-all]{960}} & \num{53653} & \textbf{\num[detect-all]{574}} \\
\num{5000.0} & \num{642088} & \num{4019} & \num{7619} & \num{856098} & \underline{\num[detect-all]{2486}} & \num{150425} & \textbf{\num[detect-all]{1278}} \\
\midrule
\multicolumn{8}{l}{\textbf{Bus Stop (761K) queried from train stations}} \\
\midrule
\num{500.0} & \num{626257} & \num{584} & \num{590} & \num{487757} & \underline{\num[detect-all]{357}} & \num{23940} & \textbf{\num[detect-all]{226}} \\
\num{1000.0} & \num{626627} & \num{882} & \num{954} & \num{493638} & \underline{\num[detect-all]{405}} & \num{29101} & \textbf{\num[detect-all]{277}} \\
\num{2000.0} & \num{629476} & \num{1387} & \num{1630} & \num{519413} & \underline{\num[detect-all]{671}} & \num{43714} & \textbf{\num[detect-all]{401}} \\
\num{5000.0} & \num{632530} & \num{3072} & \num{5205} & \num{548443} & \underline{\num[detect-all]{1870}} & \num{107848} & \textbf{\num[detect-all]{875}} \\
\midrule
\multicolumn{8}{l}{\textbf{Restaurant (103K) queried from train stations}} \\
\midrule
\num{500.0} & \num{83112} & \num{273} & \num{245} & \num{105247} & \underline{\num[detect-all]{122}} & \num{21998} & \textbf{\num[detect-all]{118}} \\
\num{1000.0} & \num{84002} & \num{382} & \num{389} & \num{106802} & \underline{\num[detect-all]{180}} & \num{20737} & \textbf{\num[detect-all]{131}} \\
\num{2000.0} & \num{85221} & \num{575} & \num{676} & \num{107534} & \underline{\num[detect-all]{255}} & \num{29747} & \textbf{\num[detect-all]{160}} \\
\num{5000.0} & \num{87189} & \num{1162} & \num{1670} & \num{118963} & \underline{\num[detect-all]{495}} & \num{37417} & \textbf{\num[detect-all]{261}} \\
\bottomrule
\end{tabular}
\end{table} 

\section{Comparison of SNN Implementations}\label{sec:snn_comparison}
Throughout the paper we used the Python implementation of SNN provided by the original paper~\cite{chenFastExactFixedradius2024}. In \cref{tab:snn_comparison} we compare this implementation to our own Rust implementation of SNN, which utilizes some 
of the low-level optimizations described in \cref{sec:perf_engineering} such as the accumulation register optimization from \cref{fig:acc_microbench}. We observe that our implementation is significantly faster than the original Python implementation for low dimensions but is beaten by the original implementation in very high dimensions. This is likely due to the original implementation using a highly optimized BLAS library for the matrix-vector multiplications. Additionally, our implementation always processes at least \num{8} points per iteration, while the vectorized approach of the original can be more efficient when the candidate set is very small.

\begin{table}[h]
	\centering
	\caption{Comparison of SNN implementations. Py-SNN is the Python implementation from the original paper \cite{chenFastExactFixedradius2024}. Our implementation "our SNN" uses optimizations from \cref{sec:perf_engineering}. \textbf{best} and \underline{second best} performing methods are highlighted.}\label{tab:snn_comparison}
	\sisetup{detect-weight=true, detect-family=true}
\begin{tabular}{l@{\,$\,$}rr@{\,$\,$}r@{\,$\,$}r@{\,$\,$}r@{\,$\,$}r@{\,$\,$}r@{\,$\,$}}
\toprule
Data Set & \multicolumn{1}{c}{$n$} & \multicolumn{1}{c}{$d$} & \multicolumn{1}{r}{\text{Brute-Force}} & \multicolumn{1}{r}{\text{SNN}} & \multicolumn{1}{r}{\text{our SNN}} & \multicolumn{1}{r}{\text{SPRK}} \\
\midrule
\multicolumn{5}{l}{\textbf{Real-World high-dimensional datasets (in \qty{}{\milli\second}) \cite{aumullerANNBenchmarks2020}}} \\
\midrule
SIFT1M  & \SI{100}{k} & \num{128} & \num{5.8} & \underline{\num[detect-all]{2.1}} & \textbf{\num[detect-all]{1.9}} & \num{2.3} \\
SIFT10K  & \SI{25}{k} & \num{128} & \num{0.9} & \textbf{\num[detect-all]{0.3}} & \textbf{\num[detect-all]{0.3}} & \num{0.4} \\
GIST  & \SI{1}{M} & \num{960} & \num{433.8} & \textbf{\num[detect-all]{183.2}} & \underline{\num[detect-all]{223.0}} & \num{255.4} \\
GloVe100  & \SI{1}{M} & \num{100} & \num{55.4} & \num{29.7} & \underline{\num[detect-all]{25.3}} & \textbf{\num[detect-all]{4.4}} \\
Deep1B  & \SI{10}{M} & \num{96} & \textbf{\num[detect-all]{308.0}} & \num{453.7} & \num{345.9} & \underline{\num[detect-all]{312.2}} \\
F-MNIST  & \SI{60}{k} & \num{784} & \num{21.2} & \textbf{\num[detect-all]{5.9}} & \num{6.4} & \underline{\num[detect-all]{6.0}} \\
\midrule
\multicolumn{5}{l}{\textbf{Clustering (in \qty{}{\nano\second}) \cite{duaUCI2017}}} \\
\midrule
Banknote  & \SI{1.5}{k} & \num{4} & \num{1750.6} & \num{21593.0} & \underline{\num[detect-all]{208.4}} & \textbf{\num[detect-all]{99.8}} \\
Wine  & \SI{0.2}{k} & \num{13} & \num{1131.4} & \num{18907.0} & \textbf{\num[detect-all]{118.6}} & \underline{\num[detect-all]{143.0}} \\
Dermatology  & \SI{0.4}{k} & \num{34} & \num{3023.2} & \num{23220.4} & \textbf{\num[detect-all]{842.4}} & \underline{\num[detect-all]{1136.4}} \\
Ecoli  & \SI{0.3}{k} & \num{7} & \num{884.4} & \num{18502.8} & \underline{\num[detect-all]{84.6}} & \textbf{\num[detect-all]{78.6}} \\
\midrule
\multicolumn{5}{l}{\textbf{Point of Interest Search (in \qty{}{\nano\second}) \cite{OsmGermany}}} \\
\midrule
ATM  & \SI{12}{k} & \num{2} & \num{10189.0} & \num{20069.5} & \underline{\num[detect-all]{115.5}} & \textbf{\num[detect-all]{86.0}} \\
Bakery  & \SI{33}{k} & \num{2} & \num{27424.0} & \num{23148.8} & \underline{\num[detect-all]{319.5}} & \textbf{\num[detect-all]{112.0}} \\
Parking  & \SI{769}{k} & \num{2} & \num{637569.5} & \num{66871.8} & \underline{\num[detect-all]{5001.5}} & \textbf{\num[detect-all]{619.5}} \\
Bus Stop  & \SI{761}{k} & \num{2} & \num{628722.5} & \num{51150.8} & \underline{\num[detect-all]{4183.5}} & \textbf{\num[detect-all]{444.8}} \\
Restaurant  & \SI{103}{k} & \num{2} & \num{84881.0} & \num{27474.8} & \underline{\num[detect-all]{598.8}} & \textbf{\num[detect-all]{167.5}} \\
\midrule
\multicolumn{5}{l}{\textbf{Embedding (in \qty{}{\micro\second})}} \\
\midrule
Embedding  & \SI{10.0}{k} & \num{2} & \num{13.1} & \num{39.2} & \underline{\num[detect-all]{1.0}} & \textbf{\num[detect-all]{0.2}} \\
Embedding  & \SI{100.0}{k} & \num{2} & \num{267.9} & \num{69.2} & \underline{\num[detect-all]{3.3}} & \textbf{\num[detect-all]{0.5}} \\
Embedding  & \SI{1000.0}{k} & \num{2} & \num{6133.1} & \num{238.8} & \underline{\num[detect-all]{17.4}} & \textbf{\num[detect-all]{2.6}} \\
Embedding  & \SI{10.0}{k} & \num{8} & \num{21.8} & \num{39.6} & \underline{\num[detect-all]{4.0}} & \textbf{\num[detect-all]{1.8}} \\
Embedding  & \SI{100.0}{k} & \num{8} & \num{398.8} & \num{227.3} & \underline{\num[detect-all]{63.4}} & \textbf{\num[detect-all]{6.2}} \\
Embedding  & \SI{1000.0}{k} & \num{8} & \num{7111.5} & \num{2174.0} & \underline{\num[detect-all]{745.4}} & \textbf{\num[detect-all]{31.1}} \\
Embedding  & \SI{10.0}{k} & \num{16} & \num{39.0} & \num{62.2} & \underline{\num[detect-all]{9.6}} & \textbf{\num[detect-all]{7.8}} \\
Embedding  & \SI{100.0}{k} & \num{16} & \num{570.7} & \num{451.5} & \underline{\num[detect-all]{216.7}} & \textbf{\num[detect-all]{141.7}} \\
Embedding  & \SI{1000.0}{k} & \num{16} & \num{11808.4} & \num{6875.8} & \underline{\num[detect-all]{3704.7}} & \textbf{\num[detect-all]{1056.5}} \\
\bottomrule
\end{tabular}
\end{table} 

\end{document}